\documentclass{statsoc}

\usepackage[T1]{fontenc}
\usepackage{amsmath}
\usepackage{amssymb}
\usepackage{tabularx}
\usepackage[dvips]{epsfig}
\usepackage[dvips]{graphics}
\usepackage{graphicx}
\usepackage{dsfont}
\usepackage{natbib}

\newcommand{\Inte}{\mathds{N}}

\newcommand{\Real}{\mathds{R}}

\newcommand{\Esp}{\mathds{E}}
\newcommand{\Var}{\mbox{Var}}

\newcommand{\Loi}{\mathcal{L}}

\def\argmin{\renewcommand{\arraystretch}{0.5}
\begin{array}[t]{c}
\text{Argmin} \\
{^{0<\tau _{1}<\tau _{2}<...<\tau _{K-1}<n}}
\end{array}\renewcommand{\arraystretch}{1}}
\def\argmini{\renewcommand{\arraystretch}{0.5}
\begin{array}[t]{c}
\text{Argmin} \\
{^{\theta \in \Theta}}
\end{array}\renewcommand{\arraystretch}{1}}

\title[A new stochastic process to model HR series and an estimator of its fractality parameter]{A new stochastic process to model Heart Rate series during exhaustive run and an estimator of its fractality parameter}
\author{Jean-Marc Bardet}
\address{Universit\'e Paris 1, SAMOS-MATISSE-CES, 90 rue de Tolbiac, 75013 Paris, France.}
\email{bardet@univ-paris1.fr}
\author{V\'eronique Billat}
\address{Universit\'e d'Evry, LEPHE, E.A. 3872 Genopole, Boulevard F. Mitterrand, Evry Cedex, France.}
\author[Bardet {\it et al.}]{Imen Kammoun}
\address{Universit\'e Paris 1, SAMOS-MATISSE-CES, 90 rue de Tolbiac, 75013 Paris, France.}
\begin{document}
\maketitle
\begin{abstract}
In order to interpret and explain the physiological signal
behaviors, it can be interesting to find some constants among the
fluctuations of these data during all the effort or during different
stages of the race (which can be detected using a change points
detection method). Several recent papers have proposed the
long-range dependence (Hurst) parameter as such a constant. However,
their results induce two main problems. Firstly, DFA method is
usually applied for estimating this parameter. Clearly, such a
method does not provide the most efficient estimator and moreover it
is not at all robust even in the case of smooth trends. Secondly,
this method often gives estimated Hurst parameters larger than $1$,
which is the larger possible value for long memory stationary
processes. In this article we propose solutions for both these
problems and we define a new model allowing such estimated
parameters. On the one hand, a wavelet-base estimator is applied to
data. Such an estimator provides optimal convergence rates in a
semiparametric context and can be used for smoothly trended
processes. On the other hand, a new semiparametric model so-called
locally fractional Gaussian noise is introduced and is characterized
by a so-called parameter which can be larger than $1$. Such
semiparametric process is tested to be relevant for modeling HR data
in the three characteristic phases of the race. It also shows an
evolution of the local fractality parameter during the race
confirming the results obtained by Peng {\em et al.} (1995) in their
study regarding Hurst parameter of HR time series during the
exercise for healthy adults (where the estimated parameter is close
to that observed in the race beginning) and heart failure adults
(where the estimated parameter is close to that observed in the end
of race). So, this evolution, which can not be observed with DFA
method, may be associated with fatigue appearing during the last
phase of the marathon.
\end{abstract}
%\end{frontmatter}
\section{Introduction}
\begin{center}
\begin{figure}[h]
\begin{center}
\includegraphics[width=9 cm,height=5 cm]{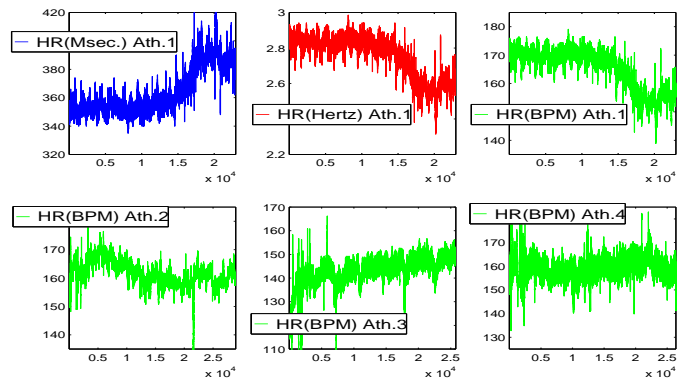}
\end{center}
\caption{Heat rate signals of Athlete 1 in ms, Hertz and BPM (up),
of Athletes 2, 3 and 4 in BPM (down)} \label{Figure1}
\end{figure}
\end{center}

The content of this article was motivated by a general study of
physiological signals of runners recorded during endurance races as
marathons. More precisely, after different signal procedures for
"cleaning" data, one considers the time series resulting of the
evolution of heart rate (HR) data during the race. The following
figure provides several examples of such data (recorded during
Marathon of Paris). For each runner, the periods (in ms) between the
successive pulsations (see Fig. \ref{Figure1}) are recorded. The HR
signal in number of beats per minute (bpm) is then
deduced (the HR average for the whole sample is of 162 bpm).\\
~\\
This paper, focuses on the modeling and the estimation of relevant
parameters characterizing these instantaneous heart rate signals of
athletes recorded during the marathon. We have chosen to focus in an
exponent that can be called "Fractal", which indicates the local
regularity of the path and the dependency between data. In certain
stationary cases, this parameter is close to the Hurst parameter,
defined for long range dependent (LRD) processes.
The LRD behavior is often seen on various data. This phenomenon was
developed in many fields beginning with hydrology (Hurst, 1951),
telecommunication, biomechanics and recently in economy and finance.
A mathematical and signal processing methods have also noted the
presence of LRD in time series describing the fluctuations over time
of physiological signals (\cite{gold1}, \cite{gold}). Indeed,
numerous authors have studied heartbeat time series (see for
instance \cite{pengm}, (1995) or \cite{asbg}) and the model proposed
to fit these data is a trended long memory process with an estimated
Hurst parameter close to $1$ (and sometimes more than $1$). In this
article, three improvements have been proposed to such a model:
\begin{enumerate}
\item Data are stepped in three different stages which are
detected using a change points detection method (see for instance
\cite{lavielle}) developed in Section 2. The main idea of the
detection's method is to consider that the signal distribution
depends on a vector of unknown characteristic parameters constituted
by the mean and the variance. The different stages (beginning,
middle and end of the race) and therefore the different vectors of
parameters, which change at two unknown instants, are estimated.
This first step is important since the model defined below fits well
for each sub-series but not at all for the whole HR data.
\item The parameter $H$ which is very interesting for
interpreting and explaining the physiological signal behaviors, is
estimating using two methods the DFA method and wavelet analysis in Section 3.\\
The DFA which is a version, for time series with trend, of the
aggregated variance method was studied in details in \cite{bardetk}.
In particular, the asymptotic properties of the DFA function and the
deduced estimator of $H$ are studied in the case of fractional
Gaussian noise and extended to a general class of stationary
semiparametric long-range dependent processes with or without trend.
We have shown that DFA method is not as efficient to estimate the
Hurst parameter of stationary long memory processes, than other
methods such as log-periodogram (see for instance, Moulines and
Soulier (2003) or wavelet analysis (see Abry {\it et al.} (1998), or
Moulines {\it et al.} (2007)), which provide the optimal convergence
rate (in sense of the minimax criterium). Moreover, if the DFA
method is not at all robust in the case of polynomial trends, this
is not such a case of the wavelet analysis method. Finally, a
goodness-of-fit test can be deduce from the wavelet analysis method.
Despite the popularity of DFA method in numerous papers concerning
such physiological signals (see for instance, Absil {\it et al.}
(1999), Ivanov {\it et al.} (2001), Peng {\it et al.} (1993),
(1995)), it is therefore clearly more interesting to
use the wavelet based estimator in view of estimate a ``fractal'' exponent of HR data. \\
As a first model, the usual fractional Gaussian noise (FGN) is then
proposed for modeling HR data. In such a context, the wavelet based
estimator provides two results. Firstly, the estimated parameter
often exceeds the value $1$, which is the largest possible value for
a FGN. Secondly, even for the $3$ different stages of the race, the
goodness-of-fit test is always rejected.
\item In Section 4, we propose to model HR data, during each stage, with a
generalization of fractional Gaussian noise, called locally
fractional Gaussian noise. Such stationary process is built from a
parameter called local fractality parameter which is a kind of Hurst
parameter in a restricted band frequency (that may take values in
$\Real$ and not only in $(0,1)$ as usual Hurst parameter). The
estimation of local fractality parameter and also the construction
of goodness-of-fit test can be made with wavelet analysis. We also
show the relevance of model and an evolution of the parameter during
the race, which confirms results obtained by other authors in their
study regarding the distinguish of healthy from pathologic data (see
Peng {\em et al.} (1995)).
\end{enumerate}
\section{Abrupt change detection} \label{sign_proc}
During effort, one or more phases can be observed in recorded HR
series, which evolve and change differently from an athlete to
another: the transition step, recorded between the race beginning
and the stage of HR reached during the effort, the main stage during
the exercise and an arrival phase until the race end. So, after
cleaning the HR data (implying that only $9$ athlete HR data are now
considered) an automatic detection of changes is applied to HR time
series cutting its in different race phases - beginning, middle and
end.\\
~\\
The change point detection method used here is developed by Lavielle
(see for instance \cite{lavielle}). The main idea is to consider
that the signal distribution depends on a vector of unknown
characteristic parameters in each stage. The different stages and
therefore the different vectors of parameters, change at unknown
instants. For instance and it will be our choice, changes in mean
and variance can be detected. Applied to the data, the change point
detection method along these two phenomena distinguishes beginning,
middle and end of race. So, it may be possible to envisage a
piecewise stationarity \emph{i.e.} that the process is almost
stationary on fixed time intervals and it remains the modeling of
this stationary component.
\subsection*{General principle of the method of change detection}
Assume that a sample of a time series $(Y(i), i=1,\ldots,n)$ is
observed. Assume also that it exists $\tau =(\tau _{1},\tau
_{2},\ldots,\tau _{K-1})$ with $0=\tau _{0}<\tau _{1}<\tau
_{2}<...<\tau _{K-1}<n=\tau _{K}$ and such that for each $j \in
\{1,2,\ldots,K\}$, the distribution law of $Y(i)$ is depending on a
parameter $\theta_j \in \Theta \subset \Real^d$ (with $d\in \Inte$)
for all $\tau_{j-1}< i \leq \tau_j$. Therefore, $K$ is the number of
segments to be deduced starting from the series and $\tau =(\tau
_{1},\tau_{2},\ldots,\tau _{K-1})$ is the ordered change instants.
Now, define a contrast function
$$
U_\theta \big (Y(\tau_j+1),Y(\tau_j+2),\ldots,Y(\tau_{j+1})\big ),
$$
of $\theta \in \Real^d$ applied on each vector $\big
(Y(\tau_j+1),Y(\tau_j+2),\ldots,Y(\tau_{j+1})\big )$ for all $j \in
\{0,2,\ldots,K-1\}$. A general example of such a contrast function
is
$$
U_\theta \big (Y(\tau_j+1),Y(\tau_j+2),\ldots,Y(\tau_{j+1})\big
)=-2\log L_\theta \big (
Y(\tau_j+1),Y(\tau_j+2),\ldots,Y(\tau_{j+1}) \big),
$$
where $L_\theta$ is the likelihood. Then, for all $j \in
\{0,2,\ldots,K-1\}$, define:
$$
\widehat \theta_j=\argmini U_\theta \big
(Y(\tau_j+1),Y(\tau_j+2),\ldots,Y(\tau_{j+1})\big ).
$$
Now, set:
$$
\widehat G(\tau_1,\ldots,\tau_{K-1})=\sum_{j=0}^{K-1}U_{\widehat
\theta_j} \big (Y(\tau_j+1),Y(\tau_j+2),\ldots,Y(\tau_{j+1})\big )
$$
As a consequence, an estimator $(\widehat \tau_1,\ldots,\widehat
\tau_{K-1})$ can be defined as:
\begin{equation}\label{estim_tau}
(\widehat \tau_1,\ldots,\widehat \tau_{K-1})=\argmin \widehat
G(\tau_1,\ldots,\tau_{K-1}).
\end{equation}
The principle of such method of estimation is very general (it can
be also devoted to estimate abrupt change in polynomial trends) and
different asymptotic behavior of the estimator $(\widehat
\tau_1,\ldots,\widehat \tau_{K-1})$ can be deduced under general
assumption on the time series $Y$ (see for instance Bai Perron,
1998,
Lavielle, 1999, or Moulines and Lavielle, 2000). \\
For HR data, it is obvious that the beginning and the end of the
race implies respectively an increasing (respectively decreasing) of
the mean of HR. However, for avoiding all confusion linked for
instance to the stress of the runner or other harmful noises, it was
chosen to detect a change in mean and variance.
\subsubsection*{Change detection in mean and variance}
\begin{center}
\begin{figure}[h]
\begin{center}
\includegraphics[width=9 cm,height=5.5 cm]{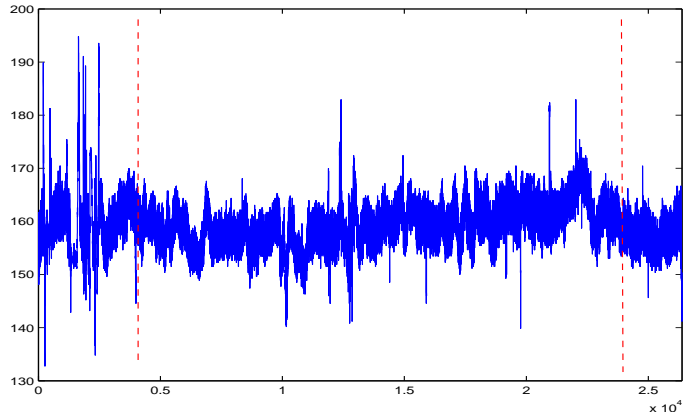}
\caption{The estimated configuration of changes in a HR time series
of an athlete} \label{Figure5}
\end{center}
\end{figure}
\end{center}

Therefore, for all $j \in \{0,1,\ldots,K-1\}$, consider the
following general model:
\[
Y(i)=\mu_{j}+\sigma _{j}\varepsilon _{i}~~\mbox{for all}~i\in
\{\tau_j+1,\ldots,\tau_{j+1}\},
\]
where $\theta_j=(m_j,\sigma_j)\in \Real \times (0,\infty)$ and
$(\varepsilon _{i})$ is a sequence of zero-mean random
variables with unit variance.\\
In the case of changes in both mean and variance, and it is such a
framework we consider for the heart rates series, a "natural"
contrast function is defined by:
\[
U_{\theta_j}\big (Y(\tau _{j}+1),\ldots,Y(\tau
_{j+1})\big)=\sum_{\ell=\tau_j+1}^{\tau_{j+1}} \frac
{(Y(\ell)-m_j)^2}{\sigma_j^2},
\]
an therefore the well-known estimator of $\theta_j$ is:
$$
\widehat \theta_j=(\widehat m_j,\widehat \sigma_j)=\Big ( \frac 1
{\tau_{j+1}-\tau_j} \sum_{\ell=\tau_j+1}^{\tau_{j+1}} Y(\ell),\frac
1 {\tau_{j+1}-\tau_j} \sum_{\ell=\tau_j+1}^{\tau_{j+1}}\big (
Y(\ell)-\widehat m_j \big )^2 \Big ).
$$
Now, the estimator $(\widehat \tau_1,\ldots,\widehat \tau_{K-1})$
can be deduced from (\ref{estim_tau}). When the number of changes is
unknown, the procedure is exactly the same except that the number of
changes is estimated. Thus, a new contrast $V$ is built by adding to
the previous contrast $U$ an increasing function depending on the
change number $K$, {\it i.e.} more precisely,
$$
\widehat V(\tau_1,\ldots,\tau_{K-1},K)=\widehat
G(\tau_1,\ldots,\tau_{K-1}) + \beta \times \mbox{pen}(K),
$$
with $\beta>0$. As a consequence, by minimizing $V$ in
$\tau_1,\ldots,\tau_{K-1},K$, an estimator
$\widehat K$ is obtained which varies with the penalization parameter $\beta$. \\
For HR data, the choice of $\mbox{pen}(K)$ was $K$. Let $\widehat
G_K=\widehat G(\widehat\tau_1,\ldots,\widehat\tau_{K-1})$, for
$K=K_1,\ldots,K_{MAX}$ we define
\[
\beta_i=\frac{\widehat G_{K_i}-\widehat G_{K_{i+1}}}{K_{i+1}-K_i}
~~\mbox{and}~l_i=\beta_i - \beta_{i+1} ~~\mbox{with}~i\geq1.
\]
Then the retained $K$ is the greatest value of $K_i$ such that $l_i>>l_j$ for $j>i$.\\
Applied to the whole set HR data, the number of abrupt changes is
estimated at 4 or 3. Three phases were selected to be studied, which
are located in the beginning of the race, in the middle and in the
end (see for example Fig. \ref{Figure5}). However for certain
recorded signals the first or the last phase can not be
distinguished probably for measurements reasons.
In order to unveil if a change of behavior of HR series was happened
during these three detected phases of the marathon, we propose a new
model for these sub-series characterized by a parameter $H$. Several
common estimators of this parameter, so-called scaling behavior
exponents, consist in performing a linear regression fit of a
scale-dependent quantity versus the scale in a logarithmic
representation. This includes the Detrended Fluctuation Analysis
(DFA) method \cite{peng} and the wavelet analysis method
\cite{abry}.
\section{A fractional Gaussian noise for modeling HR series and the
estimation of the Hurst parameter} \label{model} In this section, a
first model, the fractional Gaussian noise, is proposed for modeling
HR data. After a statistic study, one chooses to estimate the Hurst
parameter with a wavelet based estimator instead of the DFA method
(which is commonly used in physiologic papers despite its weak
performances). Moreover, a test built from the wavelet based method
shows the badness-of-fit of this model to the data.
%Both those model present are long-range dependence
%(LRD) processes, so called Hurst parameter, of the original
%signal, or the self-similarity parameter of the aggregated signal
%could be a new way for deducing some explanations.
\subsection{A first model: the fractional Gaussian noise}
\begin{center}
\begin{figure}[h]
\begin{center}
\includegraphics[width=9 cm,height=5.5 cm]{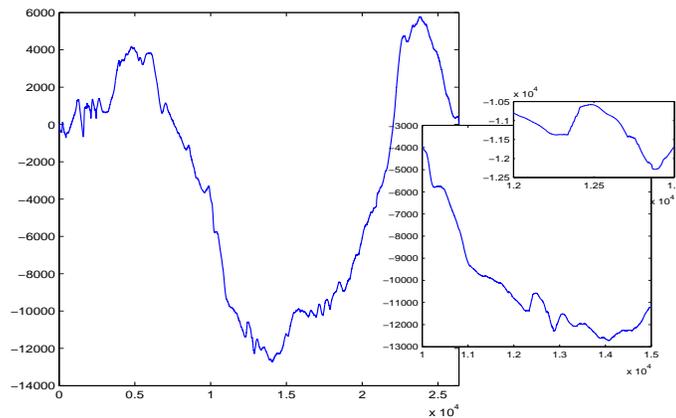}
\caption{The self-similarity of the aggregated HR signals
(representation of the aggregated HR fluctuations at 3 different
time resolutions)} \label{self1}
\end{center}
\end{figure}
\end{center}

When we observe entire or partial (during the three phases) HR time
series, we remark that it exhibits a certain persistence and the
related correlations decays very slowly with time what characterizes
trajectories of a long memory Gaussian noise. Also, the distribution
of data recorded during the phases leads as to suspect a Gaussian
behavior in these data. Of course this is only an assumption and we
can check it with tests considered for long range dependent
processes. But in our case we will try to test whether a Gaussian
process could model these data. Moreover, the aggregated signals
(see for example Fig. \ref{self1}) present a certain regularity very
close to that of fractional Brownian motion simulated trajectories
with a parameter close to 1 (Fig. \ref{Figure7}). So, fractional
Gaussian noise could be an appropriated model to HR series.
The following Figure \ref{Figure6} presents a comparison between the
graphs of HR data during a stage (detected previously) and a
fractional Gaussian noise (FGN in the sequel) with parameter
$H=0.99$ (see the definition above). Before using statistical tools
for testing the similarities of both these graphs, let us remind
some elements concerning the FGN.
\begin{center}
\begin{figure}[h]
\begin{center}
\includegraphics[width=9 cm,height=5.5 cm]{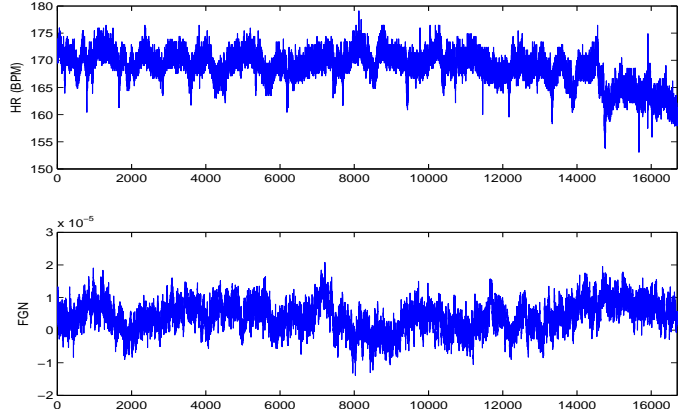}
\caption{Comparison of HR data in the middle of race (Ath4) and
generated FGN(H=0.99) trajectories} \label{Figure6}
\end{center}
\end{figure}
\end{center}
The FGN is one of the most famous example of stationary long range
dependent (LRD in the sequel) process. The LRD phenomenon was
observed in many fields including telecommunication, hydrology,
biomechanic, economy... A stationary second order process $Y=\{
Y(k),\,k \in \Inte\}$ is said to be a LRD process if:
$$
\sum_{k\in \Inte} |r_Y(k)|=\infty~~\mbox{with}~~r_Y(k)=\Esp \big [
Y(0)Y(k) \big ].
$$
Thus $Y(k)$ is depending on $Y(0)$ even if $k$ is a very large lag.
Another way for writing the LRD property is the following:
\[
r_Y(k)\sim k^{2H-2} L(k) \text{ ,   as } k \rightarrow \infty,
\]
with $L(k)$ a slowly varying function ({\it i.e.} $\forall t >0$,
$L(xt)/L(x) \to 1$ when $x \to \infty$) and the Hurst parameter
$H\in(\frac{1}{2},1)$.\\
The LRD is closely related to the self-similarity concept. A process
$X=\{X(t),\, t \geq 0\}$ is so called a self-similar process with
self-similarity exponent $H$, if $\forall c>0$:
\[
\big (X(ct)\big )_t\stackrel{\Loi}{=}c^{H} \big (X(t)\big )_t.
\]
Now, if we consider the aggregated process $\{X(t),\, t \geq 0\}$
defined by $X(k)=\sum_{i=1}^k Y(i)$ with a LRD process $Y$, then
under weak conditions (for instance $Y$ is a Gaussian or a causal
linear process), it can be proved that, roughly speaking, for $k \to
\infty$, the distribution of $\{X(t),t \geq k\}$ is a self-similar
distribution
(see Doukhan {\it et al.}, 2003, for more details).\\
The FGN is an example of a LRD Gaussian process. More precisely,
$Y^H=\{Y^H(k),\,k \in \Inte\}$ is a FGN, when
\[\label{covFGN}
r_{Y^H}(k) =\frac{\sigma ^{2}}{2}(|k+1| ^{2H}
-2|k|^{2H}+|k-1|^{2H})~~~\forall k\in \Inte,
\]
with $H\in (0,1)$ and $\sigma^2>0$. As a consequence, for $H \in
(\frac{1}{2},1)$, a usual Taylor formula implies
\[
r_{Y^{H}}(k)\sim {\sigma ^{2}}H(2H-1)k^{2H-2} \text{ ,   when  } k
\rightarrow \infty.
\]
For a zero-mean FGN, the corresponding aggregated process, denoted
here $X^H$, is so-called the fractional Brownian motion (FBM) and
$X^H$ is a self-similar Gaussian process with self-similar parameter
$H$ and therefore satisfies,
$$
\Var (X^H(k))= \sigma ^{2}|k|^{2H} \,\,\,\, \forall k\in \Inte
$$
(it can be even proved that $X^H$ is the only Gaussian self-similar
process with stationary increments). It is obvious that
$Y^H(k)=X^H(k)-X^H(k-1)$, the sequence of the increments of a FBM,
is a FGN.
\begin{center}
\begin{figure}[h]
\begin{center}
\includegraphics[width=11 cm,height=5.5 cm]{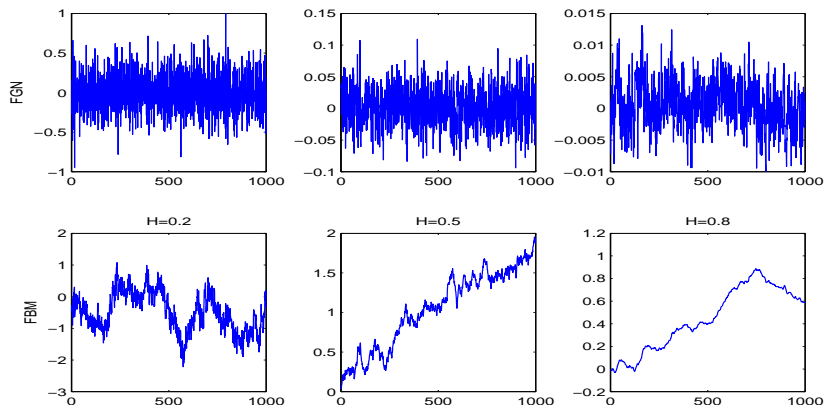}
\caption{Generated FGN trajectories and corresponding aggregated
series (FBM) for $H=0.2<0.5$ anti-persistant noise (left), $H=0.5$
white noise (center) and $H=0.8>0.5$ LRD process (right)}
\label{Figure7}
\end{center}
\end{figure}
\end{center}
Several generated trajectories of FGN and corresponding FBM are
presented in Fig. \ref{Figure7} for different values of $H$.\\
For testing if a HR path can be suitably model by a FGN, a first
step consists in estimating $H$. Here we chose to use two estimators
(but there exist many else, see for instance Doukhan {\it et al.},
2003) that are known to be unchanged to the presence of a possible
trend.
\subsection{Two estimators of the Hurst parameter: DFA and wavelet based estimators}
\begin{center}
\begin{figure}[h]
\begin{center}
\includegraphics[width=9 cm,height=6 cm]{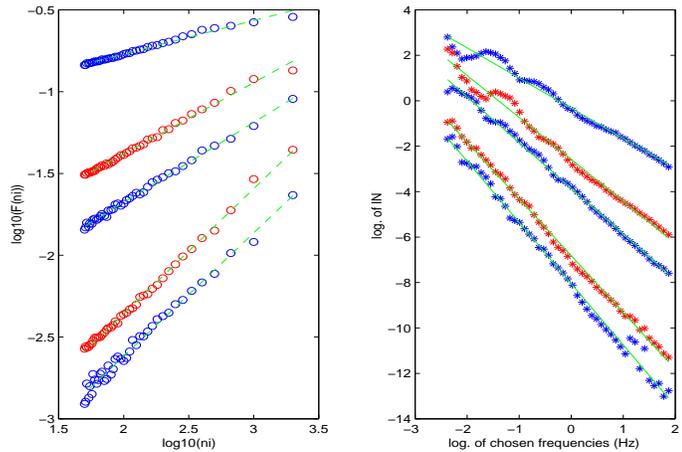}
\caption{Results of the DFA method and wavelet analysis applied to a
path of a discretized FGN for different values of $H=0.2$, $0.4$,
$0.5$, $0.7$, $0.8$, with $N=10000$} \label{Figure8}
\end{center}
\end{figure}
\end{center}

For estimating $H$, a frequently used method in the case of
physiological data processing is the Detrended Fluctuation Analysis
(DFA). The DFA method was introduced by Peng {\em et al.}
\cite{peng}. The DFA method is a version for trended time series of
the method of aggregated variance used for long-memory stationary
process. It consists briefly on:
\begin{enumerate}
\item Aggregated the process and divided it into windows with fixed
length,
\item Detrended the process from a linear regression in
each windows,
\item Computed the standard deviation of the residual
errors (the DFA function) for all data,
\item Estimated the
coefficient of the power law from a log-log regression of the DFA
function on the length of the chosen window (see Fig.
\ref{Figure8}).
\end{enumerate}
After the first stage, the process is supposed to behave like a
self-similar process with stationary increments added with a trend
(see previously). The second stage is supposed to remove the trend.
Finally, the third and fourth stages are the same than those of the
aggregated method (for zero-mean stationary process). An example of
the DFA method applied to a path of a FGN with different values of
$H$ is shown in Fig. \ref{Figure8}.
In \cite{bardetk}, the asymptotic properties of the DFA function in
case of a FGN path $(Y(1),\ldots,Y(N))$ are studied. In such a case
the estimator $\widehat H _{DFA}$ converges to $H$ with a
non-optimal convergence rate ($N^{1/3}$ instead of $N^{1/2}$ reached
for instance by maximum likelihood estimator). An extension of these
results for a general class of stationary Gaussian LRD processes is
also established. In this semiparametric frame, we have shown that
the estimator $\widehat H _{DFA}$ converges to $H$ with an optimal
convergence rate (following the minimax criteria) when an
optimal length of windows is known.\\
~\\
The processing of experimental data, and in particular physiological
data, exhibits a major problem that is the nonstationarity of the
signal. Hu {\it et al.} (2001) have studied different types of
nonstationarity associated with examples of trends and deduced their
effect on an added noise and the kind of competition who exists
between this two signals. They have also explained (2002) the
effects of three other types of nonstationarity, which are often
encountered in real data. In \cite{bardetk}, we proved that
$\widehat H _{DFA}$ does not converge to $H$ when a polynomial trend
(with degree greater or equal to $1$) or a piecewise constant trend
is added to a LRD process: the DFA method is clearly a non robust
estimation of the Hurst parameter in case of trend.\\
~\\
For improving this estimation at least for polynomial trended LRD
process, a wavelet based estimator is now considered. This method
has been introduced by Flandrin (1992) and was developed by Abry
{\em et al.} (2003) and Bardet {\em et al.} (2000). In Wesfreid {\em
et al.} (2005), a multifractal analysis of HR time series is
presented for trying to unveil their scaling law behavior using the
Wavelet Transform Modulus Maxima (WTMM) method.\\
\\
Let $\psi:\Real \rightarrow \Real$ a function so-called the mother
wavelet. Let $(a,b)\in \Real_+^*\times \Real$ and denote
$\lambda=(a,b)$. Then define the family of functions $\psi_\lambda$
by
$$\psi_{\lambda}(t)=\frac 1{\sqrt a}\,
\psi\left(\frac{t}{a}-b \right)
$$
Parameters $a$ and $b$ are so-called the scale and the shift of the
wavelet transform. Let us underline that we consider a continuous
wavelet transform. Let $d_Z(a,b)$ be the wavelet coefficient of the
process $Z=\{Z(t),\,t\in \Real\}$ for the scale $a$ and the shift
$b$, with
$$
d_Z(a,b) =\frac 1{\sqrt a} \int_{\Real}  \psi(\frac{t}{a}-b)Z(t)dt
=<\psi_{\lambda},Z>_{L^2(\Real)}.
$$
For a time series instead of a continuous time process, a Riemann
sum can replace the previous integral for providing a discretized
wavelet coefficient $e_Z(a,b)$. The function $\psi$ is supposed to
be a function such that it exists $M\in \Inte^*$ satisfying ,
\begin{eqnarray}\label{moments}
\int_{\Real}t^m \psi(t)dt = 0~~\mbox{for all $m \in \{0,1,\ldots,
M\}$}.
\end{eqnarray}
Therefore, $\psi$ has its $M$ first vanishing moments.
%%Note that it is not mandatory to choose $\psi$ to be a ``mother'' wavelet associated to a multiresolution analysis
%%of $\LL^2(\R)$ and the whole theory can be developed without resorting to this assumption~:
%Note that it is not necessary  to choose $\psi$ to be a {\em
%"mother"} wavelet associated to a multiresolution analysis of
%$\LL^2(\Real)$. The whole theory can be developed without resorting
%to this assumption. The choice of $\psi$ is then very large.\\
%\\
The wavelet based method can be applied to LRD or self-similar
processes for respectively estimating the Hurst or the
self-similarity parameter. This method is based on the following
properties: for $Z$ a stationary LRD process or a self-similar
process having stationary increments, for all $a>0$,
$(d_Z(a,b))_{b\in \Real}$ is a zero-mean stationary process and
\begin{itemize}
\item If $Z$ is a stationary LRD process,
$$
\Esp (d_Z^2(a,b))=\Var(d_Z(a,b))\sim C(\psi,H)a^{2H-1}~~\mbox{when
$a\to \infty$}
$$
\item If $Z$ is a self-similar process having stationary increments,
$$
\Esp (d_Z^2(a,b))=\Var(d_Z(a,b))\sim K(\psi,H)a^{2H+1}~~\mbox{for
all $a>0$}
$$
\end{itemize}
with $C(\psi,H)$ and $K(\psi,H)$ two positive constants depending
only on $\psi$ and $H$ (those results are proved in Flandrin, 1992,
Abry {\em et al.}, 1998). Therefore, in both these cases, the
variance of wavelet coefficients is a power law of $a$, and a
log-log regression provides an estimator of $H$. From a path
$(Z(1),\ldots,Z(N))$, the estimator will be deduced from the log-log
regression of the "natural" sample variance of discretized wavelet
coefficients, {\it i.e.},
\begin{equation}\label{SN}
S_N(a)=\frac 1 {[N/a]} \, \sum_{i=1}^{[N/a]} e_Z^2(a,i).
\end{equation}
A graph $(\log a_i,\log S_N(a_i))_{1\leq i \leq \ell}$ is drawn from
a priori family of scales and the slope of the least square
regression line provides the estimator $\widehat H_{WAV}$ of $H$. In
the semiparametric frame of a general class of stationary Gaussian
LRD processes (more general than the previous semiparametric context
required for $\widehat H_{DFA}$), it was established by Moulines
{\it et al.} (2007) that the estimator $\widehat H _{WAV}$ converges
to $H$ with an optimal convergence rate (following the minimax
criteria) when an optimal length of windows is known. Thus,
theoretical asymptotic behaviors of $\widehat H _{DFA}$ and
$\widehat H _{WAV}$ are comparable for FGN and a semiparametrical
class of LRD Gaussian processes (however more general
for $\widehat H_{WAV}$).\\
\\
This is not true any more when a polynomial trended LRD (or
self-similar) processes is considered. Indeed, Abry {\it et al.}
(1998) remarked that every degree $M$ polynomial trend is without
effects on $\widehat H _{WAV}$ since $\psi$ has its $M$ first
vanishing moments. Therefore, the larger $M$, the more robust
$\widehat H _{WAV}$ is.\\
~\\
Finally, Bardet (2002) established a chi-squared goodness-of-fit
test for a path of FBM (therefore for aggregated FGN) using wavelet
analysis. This test is based on a (penalized) distance between the
points $(\log a_i,\log S_N(a_i))_{1\leq i \leq \ell}$ and a
pseudo-generalized least square regression line (here the scales
$a_i$ are selected to behave as $N^{1/3}$).\\
~\\
In the Table \ref{Table1} appear the different estimations of $H$
computed from the DFA and wavelet analysis methods for $100$
realizations of FGN paths with $N=10000$. We choose for these
simulations the concrete procedure of wavelet analysis developed by
Abry {\it et al.} (2003) (a Daubechies wavelet is chosen and a
Mallat's fast pyramidal algorithm is used to compute wavelet
coefficients).
\begin{table}
\caption{\label{Table1}Comparison of the two samples of estimations
of $H$ with 100 realizations of fGn path (N=10000) with DFA and
wavelets methods (The \emph{p-val} was deduced from comparison of
the mean of each sample with theoretical one at the $5\%$ level)}
 \centering
\begin{tabular} {c||cccccccc}
\hline $H_{fGn}$ & $|Bias_{\widehat H_{DFA}}|$ & $|Bias_{\widehat
H_{WAV}}|$ & ${\mbox{\emph{p-val}}}_{DFA}$ &
${\mbox{\emph{p-val}}}_{WAV}$ & ${\sqrt{MSE}}_{DFA}$ & ${\sqrt{MSE}}_{WAV}$\\
\hline \hline
0.50 & 0.0064  & 0.0071 & 0.0152 & 0.0983 & 0.0271 & 0.0433 \\
0.60 & 0.0092  & 0.0009 & 0.0017 & 0.8289 & 0.0304 & 0.0405 \\
0.70 & 0.0141  & 0.0015 & $10^{-5}$ & 0.7342 & 0.0347 & 0.0436 \\
0.80 & 0.0125  & 0.0050 & 0.0002 & 0.1978 & 0.0349 & 0.0391 \\
0.90 & 0.0179  & 0.0062 & $2\cdot10^{-6}$ & 0.1030 & 0.0407 & 0.0448 \\
\hline
\end{tabular}
\end{table}

In one hand, the wavelets method appear slightly more effective than
DFA method considering the p-value which is very low for the sample
of the DFA estimations compared to wavelet analysis estimations.
This is essentially due to the estimator bias which is more
important in the case of DFA. In the other hand, if we consider the
root of MSE which is the sum of the squared bias and the variance,
the DFA estimator seems to be slightly more effective. Note that for
FGN processes (without trend), the Whittle maximum likelihood
estimator of $H$ gives a "better" results (see Taqqu {\it et al.},
1999).
\subsection{Application of both the estimators to HR data}
Both these estimators of $H$ can also be applied to the HR time
series of the $9$ athletes. The following figures Fig. \ref{Figure9}
and Fig. \ref{Figure10} exhibit examples of applications of both the
estimation method to HR data. For each athlete, it was first done to
the whole time series, and then to the different phases of the race
(as it was obtained from the detection of abrupt changes, see
Section \ref{sign_proc}). The estimation results of $H$, for the
different signals observed during the three phases of the race, are
recapitulated in the Table \ref{Table2} using wavelets method and in
Table \ref{Table3} using DFA method.
\begin{center}
\begin{figure}[h]
\begin{center}
\includegraphics[width=9 cm,height=6 cm]{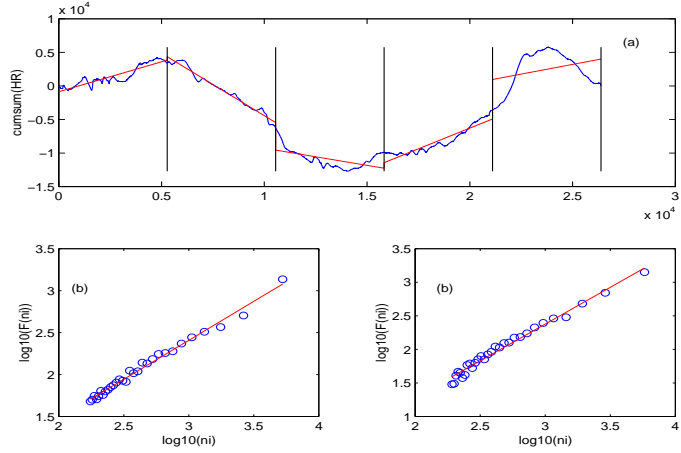}
\caption{Two first steps of the DFA method applied to a HR series
(up) and results of the DFA method applied to HR series for two
different athletes (down)} \label{Figure9}
\end{center}
\end{figure}
\end{center}
\begin{center}
\begin{figure}[h]
\begin{center}
\includegraphics[width=9 cm,height=6 cm]{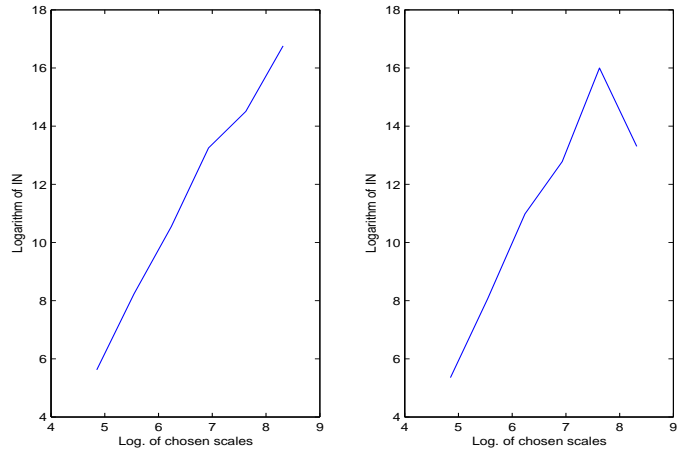}
\caption{The log-log graph of the variance of wavelet coefficients
relating to the HR series observed during the race and in the end of
race (Ath2)} \label{Figure10}
\end{center}
\end{figure}
\end{center}
\begin{center}
\begin{table}
\caption{\label{Table2}Estimated $H$ with wavelets methods for HR
series of different athletes}
\centering
\begin{tabular}{ccccccccc}
\cline{3-5} \multicolumn{1}{c}{ } & \multicolumn{1}{c}{ } &
\multicolumn{3}{c}{$Phases$}\\
\cline{2-5}
{ } & HR series & Beginning & Middle & Race end\\
\hline \hline
Ath1 & 0.8931 & 1.1268 & 1.1064 & 1.2773 \\
Ath2 & 1.1174 & 0.7871 & 1.0916 & 0.8472 \\
Ath3 & 1.0208 & 1.0315 & 1.1797 &   -    \\
Ath4 & 0.9273 &   -    & 1.0407 & 0.7925 \\
Ath5 & 1.0986 & 1.3110 & 1.0113 & 1.3952 \\
Ath6 & 1.0769 & 1.5020 & 1.1597 & 1.3673 \\
Ath7 & 1.0654 & 1.4237 & 1.1766 & 1.0151 \\
Ath8 & 0.9568 & 1.6600 & 0.9699 & 1.1948 \\
Ath9 & 0.9379 & 1.5791 & 0.9877 & 0.7263 \\
\hline
\end{tabular}
\end{table}

\end{center}
Two main problems result from these different estimations. First,
$\widehat H_{DFA}$ and $\widehat H_{WAV}$ are often larger than $1$.
However, the FGN is only defined for $H \in (0,1)$. For defining a
process allowing $H>1$, three main assumptions of FGN have to be
changed:
\begin{enumerate}
\item the assumption that the process is a stationary process;
\item the assumption that the process is a Gaussian process;
\item the assumption that only two parameters ($H$ and $\sigma^2$)
are sufficient to define the process.
\end{enumerate}
In the sequel (see below), a new model is proposed. Both the first
assumptions are still satisfied and the third one is replaced by a
semiparametric assumption.\\
~\\
The second problem is implied by the results of the goodness-of-fit
test (for wavelet analysis method). Indeed, this test is never
accepted as well for the whole time series as for the partial times
series. An explanation of such a phenomenon can be deduced from
Figure \ref{Figure10}: for the wavelet analysis, the points $(\log
a_i,\log S_N(a_i))_{1\leq i \leq \ell}$ are clearly lined for $a_i
\leq a_m$, but not exactly lined for $a_i \geq a_m$. Thus the HR
time series seems to nearly behave like a FGN for "small" scales (or
high frequencies), but not for "large" scales (or small
frequencies). A process following this
conclusion can not be the better fit of HR time series...\\
~\\
{\bf Remark:} this last conclusion leads also to a clear advantage
of wavelet based over DFA estimator. Indeed, the DFA algorithm
measures only one exponent characterizing the entire signal. Then,
this method corresponds rather to the study of "monofractal" signals
such as FGN. At the contrary, the wavelet method provides the graph
$(\log a_i,\log S_N(a_i))_{1\leq i \leq \ell}$ which can be very
interesting for analyze the multifractal behavior of data (see also
Billat {\it et al.}, 2005).
\section{A new model for modeling HR data: a locally fractional Gaussian noise}
\subsection{The locally fractional Gaussian noise}
In Bardet and Bertrand (2007), a generalization of the FBM,
so-called the $(M_K)$-multiscale FBM, was introduced. The
$(M_0)$-FBM is a FBM with self-similarity parameter $H_0$. Roughly
speaking, the $(M_K)$-FBM has the same harmonizable representation
(and therefore quite the same behavior as the FBM) than a FBM with
self-similarity parameter $H_i$ for frequencies
$|\xi|\in[\omega_i,\omega_{i+1}[$ for all $i=0,\ldots,K$
$(K\in\Inte)$. For instance, a $(M_1)$-FBM behaves as a FBM with
self-similarity parameter $H_0$ for small frequencies and as a FBM
with self-similarity parameter $H_1$ for high frequencies. Such a
model was fruitfully used for modeling biomechanical signals
(position of the center of pressure on a force platform during quiet
postural stance measured at a frequency of 100 Hz for the one minute
period).
\begin{center}
\begin{figure}[h]
\begin{center}
\includegraphics[width=9 cm,height=6 cm]{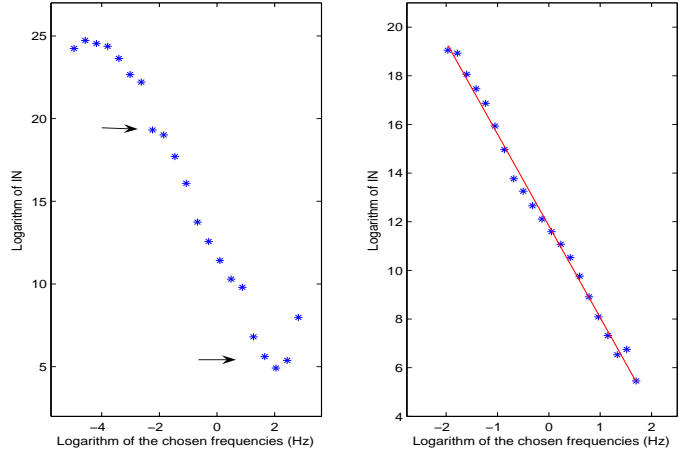}
\caption{The log-log graph of the variance of wavelet coefficients
relating to the HR series observed during the arrival phase (Ath6)
with a frequency band of [0.01 12](right) and of [0.2 4](left).}
\label{Figure11}
\end{center}
\end{figure}
\end{center}
Here, Fig. \ref{Figure11} suggests than a fitted model for
aggregated HR data should behave like a FBM with self-similarity
parameter $H$ for low frequencies and differently for high
frequencies (and not necessary like a FBM). Thus define a locally
fractional Brownian motion $X_\rho=\{X_\rho(t), t\in\Real\}$ as the
process such that:
\[
X_\rho(t)=\int_{\Real}\frac{e^{it\xi}-1}{\rho(\xi)}\widehat{W}(d\xi)
\]
where the function $\rho: \Real \to [0,\infty)$ is an even
continuous function such that:
\begin{itemize}
\item $\displaystyle \rho(\xi)= \frac 1 \sigma \, |\xi|^{H+{1/2}}~~\mbox{for
$|\xi|\in[\omega_0,\omega_1]$ with $H \in \Real$, $\sigma>0$ and
$0<\omega_0<\omega_1$}$
\item $\displaystyle \int_{\Real} \left(1\wedge  |\xi|^2\right) \frac 1 {\rho^2(\xi)}
\, d \xi < \infty.$
\end{itemize}
and $W(d\xi)$ is a Brownian measure and $\widehat{W}(d\xi)$ its
Fourier transform in the distribution meaning. Cram\'er and
Leadbetter (1967) proved the existence of such Gaussian process with
stationary increments. The main advantages of such process compared
to usual FBM are the following:
\begin{enumerate}
\item $X_\rho$ "behaves" like a FBM only for local band of
frequencies;
\item In this band, the parameter $H$ is not restricted to be in
$(0,1)$: it is in $\Real$.
\end{enumerate}
From this definition, one deduces a possible model for HR data:
\[
Y_\rho(t)=X_\rho(t+1)-X_\rho(t)=2 \cdot {\cal R}e \Big (
\int_{\Real}\frac{e^{it\xi}\sin(\xi/2)}{\rho(\xi)}\widehat{W}(d\xi)
\Big )~~\mbox{for $t \in \Real$}.
\]
Note that $Y_\rho=\{Y_\rho(t),t \in \Real\}$ is a stationary
Gaussian process and the function $2\sin(\xi/2)\rho^{-1}(\xi)$ is
so-called the spectral density of $Y_\rho$.\\
~\\
Let $\Delta_N \to 0$ and $N\Delta_N \to \infty$ when $N \to \infty$.
The wavelet based estimator can provide a convergent estimation of
$H$ when a path
$$
(Y_\rho(\Delta_N),Y_\rho(2\Delta_N),\ldots,Y_\rho(N\Delta_N))
$$
and therefore a path $(X_\rho(\Delta_N),\ldots,X_\rho(N\Delta_N))$
is observed. Indeed, consider a "mother" wavelet $\psi$ such that
$\psi:~\Real \mapsto \Real$ is a ${\cal C}^\infty$ function
satisfying~:
\begin{itemize}
\item for all $s \geq 0$, $\displaystyle{\int_{\Real} \left |t^s\psi(t)\right |dt
<\infty}$;
\item its Fourier transform $\widehat{\psi}(\xi)$ is an even function
compactly supported on  $[-\beta,-\alpha]\cup[\alpha,\beta]$ with $0
<\alpha<\beta$.
\end{itemize}
Then, using results of Bardet and Bertrand (2007), for all $a >0$
such that $\displaystyle [\frac{\alpha}{a}, \frac{\beta}{a}]
\subset[\omega_0,\omega_1]$, {\it i.e.} $\displaystyle  a \in [\frac
\beta {\omega_1}, \frac \alpha {\omega_0}]$,
$(d_{X_\rho}(a,b))_{b\in \Real}$ is a stationary Gaussian process
and
$$
\Esp \big (d_{X_\rho}^2(a,.)\big)=K(\psi,H,\sigma) \cdot a^{2H+1},
$$
with $K(\psi,H,\sigma)>0$ only depending on $\psi,H$ and $\sigma$.
However this property is checked if and only if the function $\psi$
is chosen such that:
$$
\frac \beta \alpha <  \frac {\omega_1}{\omega_0}.
$$
Moreover, for $\displaystyle  a \in [\frac \beta {\omega_1}, \frac
\alpha {\omega_0}]$, the sample variance $S_N(a)$ defined in
(\ref{SN}) and computed from a path
$(X_\rho(\Delta_N),\ldots,X_\rho(N\Delta_N))$ converges to $\Esp
\big (d_{X_\rho}^2(a,.)\big )$ and satisfies a central limit theorem
with convergence rate $\sqrt {N\Delta_N}$. Thus, with fixed scales
$(a_1,\ldots,a_\ell)  \in [\frac \beta {\omega_1}, \frac \alpha
{\omega_0}]^\ell$, a log-log-regression of $(a_i,S_N(a_i))_{1 \leq i
\leq \ell}$ provides an estimation of $H$ (and a central limit
theorem with convergence rate $N \Delta_N$ satisfied by $\widehat
H_{WAV}$ can also be established). As previously, we consider also
chi-squared goodness-of-fit test based on the wavelet analysis and
defined as a weighted distance between points $(\log ( a_i),\log
(S_N(a_i)))_{1 \leq i \leq \ell}$ and a pseudo-generalized
regression line.\\
~\\
{\bf Remark:} The main problem with these estimator and test is the
localization of the suitable frequency band $[\omega_0,\omega_1]$
%or $[\frac \beta {\omega_1}, \frac \alpha{\omega_0}]$
($\omega_0$ and $\omega_1$ are assumed to be unknown parameters). A
solution consists in selecting a very large band of scales and
determining then graphically the "most" linear part of the set of
points $(\log (a_i),\log (S_N(a_i)))_{1 \leq i \leq \ell}$. Another
possible way may be to compute an adaptive estimator of this band
using a quadratic criterion (following a similar procedure than in
Bardet and Bertrand, 2007). Here, like $9$ different paths of HR
data are observed, a common frequency band
$[\overline{\omega_0},\overline{\omega_1}]$ can be graphically
obtained and used for whole HR data (see above).
\subsection{Application to HR data}
\begin{center}
\begin{figure}[h]
\begin{center}
\includegraphics[width=9 cm,height=6 cm]{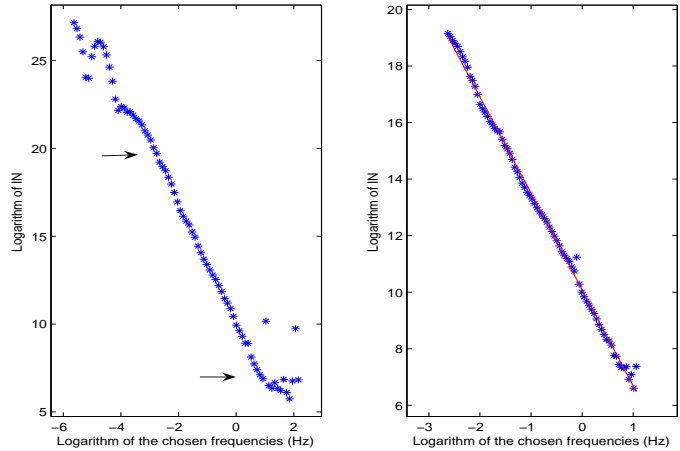}
\caption{The log-log graph of the variance of wavelet coefficients
relating to the HR series observed in the middle of the exercise
(Ath5)} \label{Figure12}
\end{center}
\end{figure}
\end{center}

\begin{center}
\begin{figure}[h]
\begin{center}
\includegraphics[width=9 cm,height=5.5 cm]{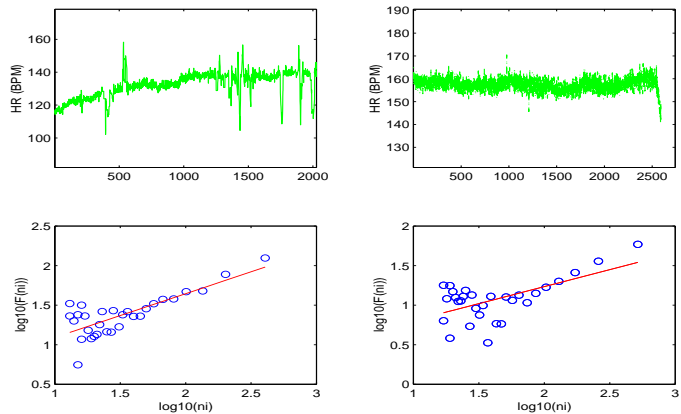}
\caption{The results of the DFA method applied to records for race
beginning (Ath3) (left) and for end of race (Ath1) (right)}
\label{Figure13}
\end{center}
\end{figure}
\end{center}

\begin{center}
\begin{figure}[h]
\begin{minipage}[c]{.45\linewidth}
\begin{center}
\begin{tabular}{c c c c c}
\hline
\multicolumn{1}{c}{ }& $\widehat H_{DFA}$ & $\widehat H_{WAV}$\\
\hline \hline
\emph{p-value} & 0.8570 & 0.0158 \\
\emph{F-stat} & 0.16 & 5.27 \\
\hline
\end{tabular}
\end{center}
\end{minipage} \hfill
\begin{minipage}[c]{.7\linewidth}
\includegraphics[width=7 cm,height=4 cm]{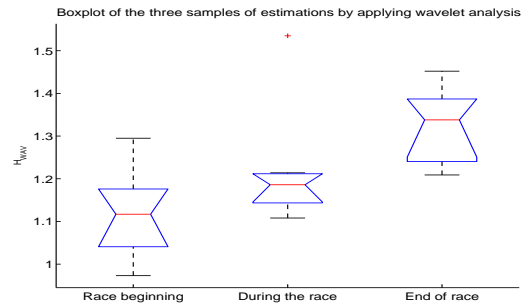}
\end{minipage}
\caption{Comparison of the three samples constituting by estimations
in the beginning of race, during the race and then in the race end
by the DFA and wavelet methods} \label{Figure14}
\end{figure}
\end{center}

First, one considers that a HR time series $(Y(1),\ldots,Y(n))$ can
be written $(Y_\rho(\Delta_n),Y_\rho(2\Delta_n),$
$\ldots,Y_\rho(n\Delta_n))$, $Y_\rho=\{Y_\rho(t),t\in \Real\}$ a
process defined as previously. Secondly, the wavelet analysis is
applied to the $9$ (whole or partial) HR time series (the chosen
"mother" wavelet is a kind of Lemarié-Meyer wavelet such that
$\beta=2\alpha$). Using first a very large band of scales for all HR
time series (for example [0.01, 12] in Fig. \ref{Figure12}), one
estimation of frequency band is deduced:
$[\overline{\omega_0},\overline{\omega_1}]=[0.2,4]$ is the chosen
frequency band for the whole and partial signals.
The estimation results of H, for the different signals observed
during the three phases of the race, are recapitulated in the Table
\ref{Table3}.
\begin{center}
\begin{table}
\caption{\label{Table3}Estimated $\widehat H$, with DFA and wavelets
methods, for HR series of different athletes $(^{*})$ The series for
which the test is rejected. Comparison of the two samples $(\widehat
H_{DFA})_{1,\ldots,9}$ and $(\widehat H_{WAV})_{1,\ldots,9}$ for
whole and partial series (p-value)} \centering
\begin{tabular}{ccccccccc}
\cline{2-9}
\multicolumn{1}{c}{ } & \multicolumn{2}{c}{HR series} & \multicolumn{2}{c}{Race beginning} & \multicolumn{2}{c}{During the race} & \multicolumn{2}{c}{End of race}\\
\cline{2-9}
\multicolumn{1}{c}{ } & $\widehat H_{DFA}$ & $\widehat H_{WAV}$ & $\widehat H_{DFA}$ & $\widehat H_{WAV}$ & $\widehat H_{DFA}$ & $\widehat H_{WAV}$ & $\widehat H_{DFA}$ & $\widehat H_{WAV}$ \\
\hline \hline
Ath1 & 0.928 & $1.288^{*}$ & 1.032 & $1.192^{~}$ & 1.060 & $1.214^{~}$ & 0.429 & $1.400^{~}$ \\
Ath2 & 1.095 & $1.268^{*}$ & 0.905 & $0.973^{~}$ & 1.126 & $1.108^{~}$ & 1.240 & $1.452^{*}$ \\
Ath3 & 1.163 & $1.048^{~}$ & 0.553 & $0.898^{~}$ & 1.130 & $1.172^{~}$ &   -   &      -      \\
Ath4 & 1.193 & $0.916^{*}$ &   -   &      -      & 1.098 & $1.249^{*}$ & 1.172 & $1.260^{~}$ \\
Ath5 & 1.239 & $1.110^{~}$ & 1.267 & $1.117^{*}$ & 1.133 & $1.205^{~}$ & 1.273 & $1.348^{*}$ \\
Ath6 & 1.247 & $1.084^{*}$ & 1.237 & $1.106^{~}$ & 1.091 & $1.172^{~}$ & 1.436 & $1.338^{~}$ \\
Ath7 & 1.155 & $1.095^{~}$ & 0.850 & $1.295^{~}$ & 1.182 & $1.186^{*}$ & 1.129 & $1.209^{~}$ \\
Ath8 & 1.258 & $1.011^{~}$ & 1.304 & $1.128^{*}$ & 0.995 & $1.134^{~}$ & 1.122 & $1.247^{~}$ \\
Ath9 & 1.243 & $1.429^{*}$ & 0.820 & $1.019^{~}$ & 1.127 & $1.535^{*}$ & 1.250 & $1.238^{*}$ \\
\hline \hline
\emph{p-value} & \multicolumn{2}{c}{$0.6414$} & \multicolumn{2}{c}{$0.3723$} & \multicolumn{2}{c}{$0.0225$} & \multicolumn{2}{c}{$0.1260$}\\
\emph{F-stat}  & \multicolumn{2}{c}{$0.23$} & \multicolumn{2}{c}{$0.85$} & \multicolumn{2}{c}{$6.38$} & \multicolumn{2}{c}{$2.65$}\\
\hline
\end{tabular}
\end{table}

\end{center}
Both DFA and wavelet analysis methods provide estimations of Hurst
exponent which reflect the possible modeling of HR data with long
range dependence time series.\\
We also note that with a p-value of $0.64$, both the samples
$(\widehat H_{DFA})_{1,\ldots,9}$ and $(\widehat
H_{WAV})_{1,\ldots,9}$
obtained from all HR time series are significantly close. \\
The same comparison can also be done when the three characteristic
stages of the race (beginning, middle and end of the race) are
distinguished. The result is different. Indeed, the corresponding
p-values between $(\widehat H_{DFA})_{1,\ldots,9}$ and $(\widehat
H_{WAV})_{1,\ldots,9}$ are significatively different in the middle
part of the race (and relatively different in the stage of race
end).\\
~\\
In spite of values relating to the estimator of $H$ for all the
athletes in the different phases which are relatively large, the DFA
has sometimes tendency to under estimating this parameter like in
the race beginning (Ath3) and the end of race (Ath1). Indeed, these
value are clearly due to a certain trend supports by the fact that
data points in log-log plot (Fig. \ref{Figure13}) have not a
straight line form, and we have proved in \cite{bardetk} that the
DFA method is not robust in the case of trended long range dependent
process. However in both the cases, the wavelets method is more
effective since it removes sufficiently this kind of trend.
For HR data and when the goodness-of-fit test is accepted, the
wavelet method shows a fractal parameter $H$ close to $1$. According
to the different studies (using DFA method) about physiologic time
series for distinguishing healthy from pathologic data sets (see
\cite{gold}, \cite{pengh}, \cite{pengm}), an exponent $H\simeq1$
indicate a healthy cardiac HR time series. Indeed, for the study
concerning a 24 hours recorded interbeat time series during the
exercise for healthy adults and heart failure adults, the following
results are obtained: for healthy subjects,
$H=1.01\pm0.16$, for the group of heart failure subjects $H=1.24\pm0.22$.\\
\\
During the different stages of the marathon race, a small increase
of the fractal parameter $H$ is observed especially at the end of
races. This behavior and this evolution may be associated with
fatigue appearing during the last phase of the marathon. This
evolution can not be observed with DFA method. Indeed, in one hand,
when we observe the three $9$-samples of wavelet estimators (related
to the $3$ phases of the race), the p-value (see Fig.
\ref{Figure14}) indicates a significantly difference due precisely
to this evolution of the fractal parameter. On the other hand, a
large p-value (0.85) is obtained for the same test using DFA
estimation.
The representation given by Fig. \ref{Figure14}, highlight a
difference in the behaviors of HR series in the beginning of the
race and in the end of race. Indeed, the dispersions in the first
and last sample are more important than in the middle of race and it
seems that each athlete starts and finished the race at his own
rhythm but in the middle athletes seems to have the same rate.
%This homogeneity of DFA estimator samples confirms the suitable
%use of this method which is to analyze the signals which
%properties do not change in the entire series i.e. monofractal
%signals. In contrast, using wavelet analysis, the fact that each
%recorded phase of race is characterized by an exponent and that
%the estimation of regularity parameter evolve during the race
%reveal the multifractal behavior of the HR time series which is
%not the case by DFA.
\section{Conclusion}
As indicated in the beginning of the last section, our main goal is
to see whether the heart rate time series during the race have
specific properties that of scaling law behavior. The wavelet
analysis and the DFA methods are applied to $9$ HR time series
during the whole and also the different three phases of the race
(beginning, middle and end of race) obtained by an automatic
procedure. Even if their results are not exactly the same, both
methods provide Hurst exponents which reflect the possible modeling
of HR data by a LRD time series. However, in \cite{bardetk}, even if
the DFA estimator of Hurst parameter is proved to be convergent with
a reasonable convergence rate for LRD stationary Gaussian processes,
it is not at all a robust method in case of trend. The wavelet based
method provides a more precise and robust estimator of the Hurst
parameter. Thus, the results
obtained from this wavelet estimator seem to be more valid. \\
~\\
Moreover, a chi-squared goodness-of-fit test can also be deduced
from this method. It seems to show that a classical LRD stationary
Gaussian process is not exactly a suitable model for HR data. Graphs
obtained with wavelet analysis also show that a locally fractional
Gaussian noise, a semiparametric process defined in Section
\ref{model} could be more relevant to model these data. A
chi-squared  test confirms the goodness-of-fit of such a model.
Thus, using the wavelet estimation of a fractal parameter in a
specific frequency band, one obtains a conclusion relatively close
to those obtained by other studies (conclusion which can not be
detected with DFA method): these fractal parameters increase through
the race phases, what may be explained with fatigue appearing during
the last phase of the marathon. Thus this fractal parameter may be a
relevant factor to detect a change during a long-distance race.\\
~\\
Finally, for the 9 athletes and as the test is validated with
significance level around 0.65, we can estimate $\widehat
H_{beginning}$ at $1.1$, the $\widehat H_{middle}$ at $1.2$ and
$\widehat H_{end}$ at $1.3$ with a larger confidence interval at the
beginning and the end of the race. This behavior could bring a new
way of understanding what is happening during a race.

%\bibliographystyle{natbib}
%\section*{Bibliography}

\end{document}